\documentstyle[twoside,11pt]{article}
\def\greaterthansquiggle{\raise.3ex\hbox{$>$\kern-.75em\lower1ex\hbox{$\sim$}}}
\def\lessthansquiggle{\raise.3ex\hbox{$<$\kern-.75em\lower1ex\hbox{$\sim$}}}
\newcommand{\beq}{\begin{equation}}
\newcommand{\eeq}{\end{equation}}
\newcommand{\beqa}{\begin{eqnarray}}
\newcommand{\eeqa}{\end{eqnarray}}
\newcommand{\ba}{\begin{array}}
\newcommand{\ea}{\end{array}}

\newcommand{\ve}{\varepsilon}
\normalsize
\begin{document}
\bibliographystyle{plain}
\begin{titlepage}
\begin{flushright}
UWThPh-1994-41\\
\today
\end{flushright}
\vspace{2cm}
\begin{center}
{\Large \bf Neutrinos Interacting with Polarizable
Media*}\\[50pt]
W. Grimus and H. Neufeld\\
Institut f\"ur Theoretische Physik\\
Universit\"at Wien\\
Boltzmanngasse 5, A-1090 Wien
\vfill
{\bf Abstract}
\end{center}

We study Cherenkov and transition
radiation of neutral spin 1/2 particles which carry magnetic
moments or electric dipole moments. In particular, we estimate
the radiation caused by the solar neutrino flux in dielectric
media.

\vfill
\noindent *Talk presented by W. Grimus at the 7th Adriatic
Meeting on Particle Physics, Brioni, Croatia, September~13--20,~1994
\end{titlepage}

\section{Introduction}
There is some hope for discovering physics beyond the Standard
Model\cite{SM} through neutrino properties like masses, mixings
and magnetic moments (MM) or electric dipole moments (EDM). The
electromagnetic interactions generated by these moments can be probed
by laboratory $\nu - e^-$ scattering yielding the bounds\cite{kr,co}
$$\mu^2_{\nu_e} + 2.1 \mu^2_{\nu_\mu} < 1.16 \cdot 10^{-18} \mu^2_B$$
\begin{flushleft} and \end{flushleft}
\beq \label{bounds}
\sqrt{\mu_{\nu_\tau}^2 + d_{\nu_\tau}^2} < 5.4 \cdot 10^{-7} \mu_B
\eeq
where $\mu_B$ is the Bohr magneton and $\mu$, $d$ denote MMs and
EDMs, respectively. Note that with the neglect of neutrino masses
MMs and EDMs cannot be distinguished and thus $\mu^2 + d^2$ is the
relevant parameter. Bounds for the $\tau$ neutrino from $e^+ e^- \rightarrow
\nu \bar \nu \gamma$ are an order of magnitude weaker\cite{gro} than
the above limit.

In this report we want to discuss the possibility of getting
information on neutrino MMs and EDMs by using their interaction
with polarizable media. It is well known that charged particles have
such interactions but also neutral ones though much weaker as long as
they have electromagnetic properties. Given a dielectric medium one
can have two opposite situations:
\begin{description}
\item[i) Cherenkov radiation] (CR)\cite{che}\\
The medium is homogeneous and infinitely extended. In this case
radiation is only possible if the velocity of the particle is larger
than the velocity of light in the medium, i.e. $v > 1/n$ where $n$ is
the refractive index.
\item[ii) Transition radiation] (TR)\cite{gif,ga,du,git}\\
There are two different homogeneous media each of which fills a
half-space. In this case radiation is emitted when the particle
crosses the boundary. There is no lower limit on $v$.
\end{description}
Quantum theoretical treatments of TR of charged particles
can be found in \cite{ga} and \cite{be}. The latter reference is
remarkable because there CR and TR for charged {\em and} neutral
particles is considered in some approximation.

In the following the quantization of the
electromagnetic field in the presence of a polarizable medium will be
discussed in
section 2. Then we will enter into CR in section 3 and consider TR in
section 4. In both cases our discussion will be generally applicable
to neutral particles with MMs and EDMs in the ultrarelativistic limit
by which we mean the limit of vanishing masses. For finite mass
effects we refer the reader to refs. \cite{gri} and \cite{gneu} for CR
and TR, respectively. In the numerical examples concerning neutrinos
we will consider the solar neutrino flux as source and normalize to a
MM of $10^{-10} \mu_B$ (\ref{bounds}). Section 5 will contain the
conclusions.

\section{Quantization of the Electromagnetic Field}
We invisage a situation where a dielectric medium is distributed in
space with its properties given by the dielectric ``constant'' $\bar
\ve (\vec x)$ varying in space but being time-independent. We
assume that the permeability of the medium is 1. Then in the absence of
free charges and currents $(\rho_f = 0,\, \vec j_f = 0)$ and with time
dependence of $\vec E$, $\vec B$ given by $e^{-i \omega t}$ Maxwell's
equations reduce to
\beq \label{max}
\mbox{ curl} \mbox{ curl} \vec E - \bar \ve \omega^2 \vec E =
\vec 0, \quad \mbox{ div} (\bar \ve \vec E)=0, \quad \vec B =
\frac{1}{i \omega} \mbox{curl} \vec E.
\eeq
This allows to dispense with the discussion of $\vec B$ most of the
time.

With $\bar \ve (\vec x) = \bar n (\vec x)^2$ we get a refractive
index varying in space. If there is some region where $\bar \ve
(\vec x) = \ve = n^2 =$ constant then a plane wave in that region
is characterized by
\beq \label{wave}
e^{-i(\omega t - \vec k \cdot \vec x)} \quad \mbox{with} \quad \vec
k^2 = n^2 \omega^2.
\eeq
Now we assume that the medium is distributed in such a way that one
can split up the fields in
\beq \label{emf}
\vec E_j(\vec k,x) \sim \vec e_j e^{-i(\omega t - \vec k \cdot \vec
x)} + \mbox{``scattered'' part}
\eeq
with $j = I, II$ indicating the polarizations. This allows to label
the solutions of (\ref{max}) by the momentum $\vec k$ of the incident
plane wave. If in addition the fields are normalized
such that
$$
\frac{1}{2} \int d^3x \left(\bar \ve(\vec x) \vec E_j(\vec k,x)^* \cdot
\vec E_{j'}(\vec k',x) + \vec B_j(\vec k,x)^* \cdot
\vec B_{j'}(\vec k',x)\right) = (2 \pi)^3 \delta_{jj'} \delta(\vec k -
\vec k')
$$
\beq \label{N}
\int d^3x \left(\bar \ve(\vec x) \vec E_j(\vec k,x) \cdot
\vec E_{j'}(\vec k',x) + \vec B_j(\vec k,x) \cdot
\vec B_{j'}(\vec k',x)\right) = 0
\eeq
then the quantization is straightforwardly performed by
\beq \label{E}
\vec E(x) = \sum_{j=I,II} \int \frac{d^3k}{(2\pi)^{3/2}}
\sqrt{\frac{\omega}{2}} \left(a_j(\vec k) \vec E_j(\vec k,x) +
a^\dagger_j(\vec k) \vec E_j(\vec k,x)^*\right)
\eeq
and
\beq
[a_j(\vec k),a^\dagger_{j'}(\vec k')] = \delta_{jj'} \delta(\vec k - \vec k'),
\qquad [a_j(\vec k), a_{j'}(\vec k')] = 0.
\eeq
For $\vec B$ a relation analoguous to (\ref{E}) holds. Then the
Hamilton operator is given by
\beq
H_\gamma = \frac{1}{2} \int d^3 x: \left(\bar \ve (\vec x) \vec E{}^2(x) +
\vec B{}^2(x)\right) : = \sum_{j=I,II} \int d^3 k \, \omega \,
a^\dagger_j(\vec k)
a_j(\vec k).
\eeq
The correctness of the quantization procedure is checked by the time
evolution
\beq
i[H_\gamma , \vec E(x)] = \dot{\vec E}(x), \quad i[H_\gamma , \vec
B(x)] = \dot{\vec B}(x).
\eeq
Given equs. (\ref{N}) this check is nearly trivial. Depending on the
physical problem it might not be necessary to find a complete system
of solutions (\ref{max},\ref{emf}). Only those electromagnetic
fields appearing in the problem are required.

In the following we want to treat two special cases of the reaction
${\cal P} \rightarrow {\cal P}' + \gamma$ in the presence of the
medium where ${\cal P}$, ${\cal P}'$ are arbitrary particles with
the same electric charge which are assumed to have tree level
electromagnetic interactions at least in some effective theory. We
are thus considering the tree graph of the above reaction but with
the photon being represented by (\ref{emf}) insted of vacuum plane
waves. With $p_i, p_f, k$ being the 4-momenta of ${\cal P}, {\cal
P}', \gamma$, respectively, and thus $p^2_i = m^2_i, p^2_f = m^2_f$,
the process ${\cal P} \rightarrow {\cal P}' + \gamma$ is only
possible in vacuum if $m_i > m_f$. The presence of the dielectric
medium, however, in general allows this reaction even for $m_f \geq
m_i$ in a certain range. Since we have assumed the medium to be static the
energy is still conserved but in general momentum is not. We can
summarize the situation in the following way:
\begin{displaymath}
q \equiv p_i - p_f \: \Rightarrow \left\{
\begin{array}{cl} \mbox{vacuum:} & q=k\\
             \mbox{medium:} & q^0 = \omega \mbox{ but } \vec q
\not= \vec k \mbox{ in general.}
\end{array} \right.
\end{displaymath}

As mentioned in the introduction we are interested in the reaction
\beq \label{rad}
\nu(p_i,s_i) \rightarrow \nu'(p_f,s_f) + \gamma_j(k)
\eeq
where $\nu, \nu'$ are neutral spin 1/2 particles with polarizations
$s_i, s_f$, respectively, and the tree level electromagnetic
interaction is given by (transition) MMs and/or EDMs, i.e.
\beq\label{emc}
\langle p_f,s_f | J^\mu_{\rm em}(x) | p_i,s_i\rangle = \frac{-i}{(2\pi)^3}
\left( \frac{m_f m_i}{E_f E_i} \right)^{1/2} e^{-iq \cdot x}
\bar u_f ( \mu + i d \gamma_5) \sigma^{\mu\nu} q_\nu u_i
\eeq
with $u_i \equiv u(p_i,s_i)$ etc.

\section{Cherenkov Radiation}
Since here the medium is assumed to be infinitely extended and
homogeneous it is the only case where one has $q=k$ and
\beq
\vec E_j(\vec k,x) = \frac{1}{n} \vec e_j e^{-i(\omega t - \vec k
\cdot \vec x)}.
\eeq
The factor $1/n$ comes from the normalization condition (\ref{N}).
The angle $\theta = < \!\!\! )(\vec p_i,\vec k)$ under which the
photon is emitted is given by
\beq\label{ca}
\cos \theta = \frac{1}{vn} \left[ 1 + (n^2 -1) \frac{\omega}{2E}
+ \frac{m^2_f - m_i^2}{2 E \omega} \right]
\eeq
where $v$ is the velocity of $\nu$ and $E \equiv E_i$. With
\beq\label{ap}
\frac{\omega}{E} \ll 1, \quad \frac{|m^2_f - m^2_i|}{E\omega} \ll 1
\eeq
it reduces to the well-known Cherenkov angle
\beq
\cos \theta = \frac{1}{vn}.
\eeq

As discussed in the previous chapter there are two cases:
\begin{displaymath}
\begin{array}{ccl}
m_f \geq m_i & \Rightarrow & n > 1, \: v > 1/n \quad \mbox{CR}\\
m_f < m_i    & \Rightarrow & \mbox{decay } \nu_i \rightarrow \nu_f +
\gamma \mbox{ in the medium.}
\end{array}
\end{displaymath}
For an upper limit on $m_f$ in the first case see \cite{gri}.
Only in the first case one has genuine CR but in the limit (\ref{ap})
both cases become indistinguishable. Since we are interested in $m_i,
m_f \rightarrow 0$ this feature is inherent in our discussion.
Therefore, for simplicity, we will only talk about CR radiation
in the following. For TR the situation is analoguous and the
distinction between the two cases will not be mentioned there
anymore.

In the limit of vanishing masses the transition rate for CR (summed
over $s_f$ and the photon polarizations) is given by the simple
formula
\beq
dR = \frac{\alpha}{4 m_e^2} \frac{\mu^2 + d^2}{\mu_B^2}
\left(n-\frac{1}{n}\right)^2 \omega^2 d\omega \label{dR}
\eeq
where $m_e$ is the electron mass and $\alpha$ the fine structure
constant. Note that $dR$ is independent of $E$ and the initial
polarization $s_i$.

Let us now apply (\ref{dR}) to the case of solar neutrinos. If $p_a$
is the probability to find $\nu_a \: (a=e,\mu,\tau)$ in the solar
neutrino flux then actually\cite{gri}
\beq
\mu^2_{\rm eff} = \sum_{a,b} (|\mu_{ba}|^2 + |d_{ba}|^2)p_a
\eeq
is probed in CR where $\mu_{ba}, d_{ba}$ are transition moments from
neutrino flavour $a$ to $b$. (Also in TR $\mu_{\rm eff}$ is the
relevant quantity.) Since we need a large volume of a medium with
$n>1$ it is obvious to take water with $n \simeq 1.335$ in the range
of visible light $\hbar \omega_1 = 1.7$ eV $ < \hbar \omega <
\hbar \omega_2 = 3$ eV (water is practically opaque in the ultraviolet
region) and with the Cherenkov angle $41,5^0$ for $v=c$. Then the number of
photons emitted by the neutrinos in the solar neutrino flux $I$
during a time intervall $T$ and in an observation volume $V$ is given
by
\beq\label{ngamma}
N_\gamma = \frac{TIV}{c} \int_{\omega_1}^{\omega_2} dR =
\frac{T}{1\mbox{day}} \, \frac{I}{6 \cdot
10^{10}\mbox{cm}^{-2}\mbox{s}^{-1}} \,\frac{V}{1\mbox{km}^3}
\,\left(\frac{\mu_{\rm eff}}{10^{-10}\mu_B}\right)^2 \cdot 46.4.
\eeq
In equ. (\ref{ngamma}) the number of photons has been normalized
to the solar neutrino flux as obtained by the solar standard
model\cite{ba}.

Finally we want to estimate CR of ultrarelativistic neutrons  with
$\mu_n \simeq -10^{-3}\mu_B$. If one has a beam with $i_n$ neutrons per
second then
\beq\label{nn}
N_\gamma \sim 10^{-15} i_n T \frac{a}{1 \mbox{\ cm}}
\eeq
where $a$ is the thickness of the medium.

Both results (\ref{ngamma},\ref{nn}) show that it is exceedingly difficult
to measure CR of neutral particles.

\section{Transition Radiation}
We consider a slab of dielectric medium situated in space between the
planes $z=-a/2$ and $z=a/2$. Thus we have two surfaces. The wave
vectors of plane wave solutions of Maxwell's equations can be
parametrized by
\beq
\vec k = \omega \left( \ba{c} \sin \alpha \cos \phi \\
\sin \alpha \sin \phi \\ \cos \alpha \ea \right) , \qquad
\vec k_m = n \omega \left( \ba{c} \sin \beta \cos \phi \\
\sin \beta \sin \phi \\ \cos \beta \ea \right)
\eeq
corresponding to vacuum and medium, respectively. In addition there
are the wave vectors $\vec k_r, \vec k_{mr}$ corresponding to
reflected waves outside and inside the dielectric slab, respectively.
By Snell's law we obtain
\beq
\sin \alpha = n \sin \beta
\eeq
and therefore $q_x = k_x = k_{mx}, \: q_y = k_y = k_{my}$ but $q_z
\not= k_z$. This reflects the translation symmetry of the problem in
the $xy$-plane. As it will turn out that for us only the photon polarization
$j=II$ in the plane of incidence is important we sketch the electric
field in this case:
\beqa\label{EII}
\vec E_{II}(\vec k,x) &=& e^{- i \omega t} \cdot \left\{
\ba{ll} \vec e_{II} e^{i \vec k \cdot \vec x} + \vec e_{IIr} a^{II}_r
e^{i \vec k_r \cdot \vec x}, & z < - a/2 \\[5pt]
\vec e_{IIm} a^{II}_m e^{i \vec k_m \cdot \vec x} +
\vec e_{IImr} a^{II}_{mr} e^{i \vec k_{mr} \cdot \vec x}, &
- a/2 < z < a/2 \\[5pt]
\vec e_{II} a^{II}_t e^{i \vec k \cdot \vec x}, & z > a/2 . \ea \right.
\eeqa
For $\vec E_I(\vec k,x)$ and the exact definitions of
the polarization vectors and the
coefficients in $\vec E_{II}(\vec k,x)$ obtained by continuity
conditions see ref. \cite{gneu}. For
simplicity we have only discussed $k_z > 0$.

The computation of the probability $W$ for the process (\ref{rad})
is extremely simplified by the use of the relation
\beq
i(E_i - E_f) \int d^4 x A_\mu(x) \langle p_f,s_f|J_{\rm em}^\mu(x)
| p_i,s_i\rangle = \int d^4x \vec E(x) \cdot
\langle p_f,s_f | \vec J_{\rm em}(x) | p_i,s_i\rangle.
\eeq
With equ. (\ref{emc}) the expressions
\beq\label{ez}
\vec {\cal E}_j(\vec k,q_z) =
\left. \int \! dz \, e^{-iq_zz} \vec E_j(\vec k,x) \right|_{x^0 = x^1
= x^2 = 0}
\eeq
appear in the probability amplitudes by $z$-integration whereas in
$x$ and $y$-directions the $\delta$-functions express momentum
conservation. For the full expression of $d^6 W$, equ. (\ref{ez}) and
further computational details we refer the reader again to ref.
\cite{gneu}.
We now concentrate on the limit $m_i, m_f \rightarrow 0$ and on the
case where the incident particle flux is orthogonal to the dielectric
layer. One can show that this leads to the following consequences:
\begin{itemize}
\item The TR does not depend on the initial polarization $s_i$ of the
particle.
\item Only photon polarization $II$ contributes.
\item The situation is invariant under rotation around the $z$-axis
and therefore the differential probability of (\ref{rad}) is independent
of the angle $\phi$.
\end{itemize}

With $\theta$ being the angle between $\vec k$ and the $z$-axis one
obtains the rather simple result\cite{gneu}
\beqa
\label{urw}
d^2W &=& \sum_{\eta = \pm 1} \frac{\sin^3\theta d\theta \omega^3 d\omega}
{2 \pi^2} \frac{E}{P} (\mu^2 + d^2)\left( n - \frac{1}{n}\right)^2
\cdot \nonumber \\
&& \cdot \frac{1}{q^2_z - k^2_z} |S_-^{II}(q_z - \omega - k_{mz}) +
S_+^{II} (q_z - \omega + k_{mz})|^2
\eeqa
with
$$
q_z = E - \eta P, \qquad
P = (E^2 - 2E\omega + \omega^2 \cos^2\theta)^{1/2},
$$
\beq
S^j_- \equiv a^j_m \frac{\sin \frac{a}{2}(k_{mz} - q_z)}{k_{mz} - q_z},
\qquad
S^j_+ \equiv a^j_{mr} \frac{\sin \frac{a}{2} (k_{mz} + q_z)}
{k_{mz} + q_z} .
\eeq
$\eta = \pm 1$ corresponds to forward and backward scattering,
respectively, of the fermion.

Dielectric media usually become transparent for photons in the X-ray
range. There one has a simple expression for the dielectric constant,
namely $\ve = n^2 = 1 - \omega^2_p/\omega^2$ with $\omega_p$ being
the plasma frequency. In the following we will take $\omega_p = 20$
eV of polypropylene as a typical example. Furthermore, the limit
$\omega \gg \omega_p$ and a realistic thickness $a$ of the dielectric
layer leads to $a \omega \gg 1$ (e.g. $a \omega \simeq 10^6$ for
$a=0.01$ mm and $\omega = 20$ keV). It is easy to check that now only
small angles $\theta$ contribute and that the following approximations
can be made:
$$
|a^{II}_m| \rightarrow 1, \quad a^{II}_{mr} \rightarrow 0,
$$
\beq
q_z \simeq \omega + \frac{\omega^2 \theta^2}{2(E-\omega)},\quad
k_{mz}-q_z \simeq -\frac{\omega}{2} \left((\frac{\omega_p}{\omega})^2
+ \frac{E}{E-\omega} \theta^2 \right)
\eeq
and
$$
dW \simeq \frac{(\mu^2 + d^2)}{2 \pi^2} \left(
\frac{\omega_p}{\omega}\right)^4 \omega^3 d\omega \int_0^\infty  d \theta \;
\theta
\left\{ \frac{\sin \frac{a \omega}{4} [( \frac{\omega_p}{\omega})^2
+  \frac{E}{E - \omega} \theta^2 ]}
{\frac{\omega}{2} [(\frac{\omega_p}{\omega})^2 +
\frac{E}{E - \omega} \theta^2]}\right\}^2 =
$$
\beq \label{res}
= \frac{(\mu^2 + d^2)a}{4 \pi^2} d\omega \; \omega^2
\frac{E - \omega}{E} \left( \frac{\omega_p}{\omega}\right)^4
\int_{y_0}^\infty \left( \frac{\sin y}{y} \right)^2
\mbox{ with  } y_0 = \frac{1}{4} a \omega \left(
\frac{\omega_p}{\omega} \right)^2.
\eeq

To apply (\ref{res}) we consider the numerical example
$a=0.01$ mm and $\hbar \omega_1 = 20$ keV $\leq \hbar \omega \leq \hbar
\omega_2$. Then with $\omega_2 \gg \omega_1$ and $y_0 \simeq 0.25$ we
estimate the probability $W$ for the emission of a photon in the
energy range $[\omega_1, \omega_2]$ when one fermion is crossing the
slab by
\beq
W = \int_{\omega_1}^{\omega_2} dW \sim \frac{1}{8\pi} (\mu^2 + d^2)
\frac{\omega_p^4 a}{\omega_1} \sim 10^{-12} \frac{\mu^2 +
d^2}{\mu_B^2}.
\eeq

This very small probability in conjunction with the upper limit on
the neutrino MMs and EDMs cannot be overcome by the large solar
neutrino flux. Taking $\mu_{\rm eff} / \mu_B = 10^{-10}$ and $10^5$
foils with 10 m$^2$, results in an order of magnitude of $10^{-4}$
photons per year. A larger photon yield can only be expected for a
sizeable fraction of $\tau$ neutrinos in the solar neutrino flux
with $\mu_{\tau \tau}$ close to the experimental upper
limit (\ref{bounds}).

In the case of ultrarelativistic neutrons with an optimistic flux of
$10^{15}$ particles per second and $10^3$ foils one expects around
one photon per second.

\section{Conclusions}
Finally we want to present a short summary of our results and stress
some physical aspects of the considered radiation mechanism.
\begin{itemize}
\item In this report we have considered CR and TR of
ultrarelativistic neutral spin 1/2 particles with MMs and EDMs.
Because of the nature of the considered electromagnetic interaction,
a spin flip of the fermion must occur. We want to stress therefore
that only the quantum theoretical calculation gives a correct result
whereas the classical calculation gives zero in the limit of
vanishing particle masses\cite{sa}.
\item We have discussed two ranges of photon energy, visible light
and frequencies much larger than the plasma frequency. Between the
two ranges, media are usually non-transparent. Thus we can summarize
the situation in the following way.
\begin{itemize}
\item $\gamma$ in keV range: $\ve = 1 - \omega_p^2/\omega^2 < 1 \:
\Rightarrow$ only TR possible.
\item $\gamma$ in optical range: $\ve = n^2 > 1$ and finite
dielectric slabs $\Rightarrow$ The above described computation of TR
gives the exact result for radiation but
for $a \rightarrow \infty$ (i.e. $a \omega \gg 1$) formula
(\ref{urw}) develops the form $adR$ of CR\footnote{We have checked
this only semiquantitatively.}\hspace{1em}with the angle of emission $\theta'$
given by $\cos \theta'= \sqrt{2-n^2}$. This angle is obtained from the
Cherenkov angle $\cos \theta = 1/n$ by taking into account refraction
at the surface of the layer. Thus we effectively have CR radiation in
the optical range and we confirm that for $n>1$ equ. (\ref{urw}) is the
general expression for radiation.
\end{itemize}
\item CR and TR are independent of the initial polarisation $s_i$.
Therefore, also spin flipped neutrinos would be counted in the solar
neutrino flux.
\item Unfortunately, for neutrinos the photon yield is very low
because of the upper bounds on their MMs and EDMs.
\item There might be some hope to see such effects for neutrons.
\end{itemize}

\section*{Acknowledgement}
We thank the organizers for the pleasant and stimulating atmosphere.

\end{document}